\documentclass[preprint,showpacs,preprintnumbers,amsmath,amssymb,superscriptaddress,floatfix]{revtex4}

\usepackage{graphicx}
\usepackage{dcolumn}
\usepackage{bm}
\usepackage{color}
\usepackage{hyperref}

\begin{document}
\title{Single-particle Analysis of Non-interacting Ultracold Bosons in Amplitude Modulated Parabolic Optical Lattice}
\noindent
\author         {Tomotake Yamakoshi}
\affiliation{Department of Engineering Science, University of Electro-Communications, 1-5-1 Chofugaoka, Chofu-shi, Tokyo 182-8585, Japan}
\author         {Shinichi Watanabe}
\affiliation{Department of Engineering Science, University of Electro-Communications, 1-5-1 Chofugaoka, Chofu-shi, Tokyo 182-8585, Japan}

\date{\today}
\pacs{03.75.-b,03.75.Lm,37.10.Jk,67.85.-d}

\begin{abstract}
Ultracold atoms in the combined potential of a parabolic trap and an optical lattice is considered a promising tool for coherent manipulation of matter wave packets.
The recent Aarhus experiment[P.~L.~Pedersen {\it et al.}, Phys.~Rev.~A.~{\bf 88}, 023620 (2013)] produced wave packets  by applying the optical lattice's amplitude modulation to a Bose-Einstein condensate (BEC) of $^{87}$Rb.
The present paper renders a theoretical account with single-particle analysis of this experimental production of the wave packets and their subsequent time-evolution.
We focus on the one-dimensional non-interacting bosonic system as a fundamental starting point for accurate quantum analysis and for further investigation of similar experiments.  
We show that a simple Rabi-oscillation model gives a good description of the wave packet production in terms of the inter-band transition while the first-order perturbation theory proves inadequate, that is the recent experiment already reached the realm of high-order couplings.
As a natural extension, we demonstrate enhancement of the wave packet production by the two-step Rabi-oscillation method using either single frequency or dual frequencies.
We assess the high-order Bragg reflection and Landau-Zener transition at a band gap with the aid of rigorous quantum time-propagation as well as the semi-classical theory exploited earlier by the Hamburg experiment [J.~Heinze {\it et al.}, PRL~{\bf 107}, 135303(2011)].
Complicated reflections and splittings of the wave packet during free evolution may be largely attributed to the intertwining of these two effects.
\end{abstract}

\maketitle

\section{introduction}
\label{sect:Introduction}
Ultracold atoms in the optical lattice (OL) have been eagerly studied since the experimental realization of Bose-Einstein condensates\cite{BECR} and quantum degenerate Fermi gases\cite{DFGR}.
Their high controllability and visibility render the systems suitable for the investigation of quantum vortices\cite{VORTEX}, force detector\cite{GRAV}, quantum simulator\cite{SIMU}, artificial gauge field\cite{GAUGE}, and so on.
In particular, a great deal of attention has been drawn to the Bose-(Fermi-)Hubbard model\cite{HUM,BHUM}, which revealed a plethora of novel quantum phases in relatively deep OL potentials. 
In contrast, the dynamics of ultracold atoms in rather shallow OLs may be investigated by analogy to the electrons in crystals. 
The energy band structure, namely a general consequence of the periodic potential, naturally plays an important role in coherent manipulation of the matter waves. 
Pulse-like modulation of the OL amplitude allows us to selectively excite  subsets of atoms in specific quasimomentum states. 
Bloch oscillations\cite{LBO} and Landau-Zener (LZ hereafter) transitions~\cite{LZZ} were observed in the past in ultracold atomic systems in OL in the presence of a constant force.
These facts suggest that interferometric techniques for splitting and recombining the wave packets are feasible.

Recently, there is a revival of interest in atoms in the combined potential of the parabolic potential and the optical lattice (parabolic OL hereafter).
One interesting feature of the system is that some of the eigenstates are spatially localized far away from the center of the parabolic potential\cite{Ott,Single,Ana,Eigens}.
It is suggested that this kind of Spatially Localized Eigenstates (referred to SLEs) be used and controlled as quantum registers\cite{REGI} with precision. 
The recent BEC experiment done in Aarhus achieved an efficient population transfer of atoms to  localized states and a wave packet manipulation in the following way\cite{Arlt1}.
First, excite BEC by lattice amplitude modulation and thus create wave packets in a higher energy band.
And then after a few milliseconds of free propagation, de-excite the atoms by another amplitude modulation at such a timing that the final wave packet is concentrated in a spatial region where the trap and OL are energetically balanced. The following two conditions are necessary for a large population transfer to occur, be it excitation or de-excitation, namely a large spatial overlap and the energy conservation, of which the latter means the equality of the modulation frequency and the energy difference between the initial and final states (divided by $\hbar$).
The Aarhus group inferred the motion of the wave packet using a simple semiclassical method\cite{Heinze} and neglecting the atom-atom interaction.
Their estimates agreed qualitatively but not quantitatively with the experimental results.
Moreover, when the net excitation energy exceeds the height of the optical lattice, there is no satisfactory theoretical analysis. 
This work is motivated by this current situation.

We focus on the excitation followed by the free wave packet propagation since this sequence contains the gist of the Aarhus experiment.
Importantly, we use parameter values drawn from the actual experiment.
In doing so, we consider non-interacting bosonic systems in a one-dimensional parabolic OL, and reveal how the amplitude modulation pulse affects futures of the band population as a function of time. 
Most importantly, it must be stated that the spatial variation of the parabolic trap is so gentle and slow in comparison to the spatial oscillations of the OL that the trap affects the periodicity of the OL very adiabatically. 
The energy band structure thus plays a crucial role.
A most enlightening result of our investigation is that a simple Rabi model adapted to the Bloch energy band structure applies; its reliability is unexpectedly high.
This suggests that the amplitude modulation may be treated in analogy with radiation fields, thus known techniques in quantum optics are applicable except for a slight difference in selection rules. 
We shall treat the amplitude modulation of the OL as if it were the ``{\it radiation field}''.

In order to interpret our numerical results, we found it effective to extend the semiclassical theory originally applied to the wave packet motion of fermions in a parabolic OL in Ref.~\cite{Heinze}. Such theory gives a useful guideline and language for understanding the dynamics if somewhat imperfect.
In the Aarhus experiment, the excited wave packet gets accelerated under the influence of the parabolic potential. 
This fact leads, on the one hand, to the high order Bragg diffraction and, on the other, to the Landau-Zener transition at band gaps.
To digress on the second of these two effects, the energy gap between two neighboring bands generally decreases with increasing band index.
When the band gap becomes negligibly small, the LZ transition rate reaches 100\%. However, in the parameter regime of the actually conducted experiments, the LZ transition rate departs from this limiting value. 
We employ the simple semiclassical theory to estimate the high-order Bragg diffraction and the LZ transition in Sect.~\ref{ss:Bloch-LZ}.

It is well-known that the atom-atom interaction, represented as the nonlinear term in the Gross-Pitaevskii equation, plays an important role in ultracold bosonic systems. 
However, the experimental results of Aarhus\cite{Arlt1,Arlt2} show that the nonlinear term affects the major dynamics of the wave packet in a rather insignificant manner.
And, the wave packet dynamics shows very complicated features even in the linear case because the parabolic potential modifies the usual Bloch band picture.
Hence, it would make sense to analyze complexities due to nonlinearity in reference to the linear case.
The ideal single-particle treatment carried out here is intended to give a good starting point for analyzing experiments with a variable interaction strength\cite{Heinze}.
This paper thus focuses on the linear system foregoing the examination of nonlinear effects, a future endeavor.
Incidentally, the atom-atom interaction could be tuned to 0 anyhow by adjusting the Fano-Feshbach resonance position by an external magnetic field\cite{Chin}.


The paper is organized as follows.
Sect.~\ref{sect:system} outlines the system and its theoretical model, and gives some additional background about the experiment. 
Sect.~\ref{sect:numerics} discusses the numerical results of the population transfer with single and two-color excitation.
Sect.~\ref{sect:subsequent} discusses the analysis of the dynamics of the wave packet after the excitation process.
Sect.~\ref{sect:conclusions} concludes the paper.
This paper uses recoil energy $E_r=\hbar^2 k_{r}^2/2 m$ for the unit of energy, recoil momentum $k_{r}=2\pi/\lambda$ for the unit of (quasi-)momentum, lattice constant $a=2/\lambda$ for the unit of length and rescaled time $t=E_r t'/\hbar$ for the unit of time.
Here $\hbar$, $\lambda$ and $m$ correspond to the Planck constant, wave length of the optical lattice, and mass of the particle, respectively. 

\section{System in combined potential of parabolic and optical lattice}
\label{sect:system}
Experimental creation and detection of wave packets\cite{Arlt1,Arlt2,Heinze} consists principally of three independent manipulations, namely initial state preparation, excitation or deexcitation with amplitude modulation, and free propagation on top of either band-mapping or {\it in situ} observation.
The time-dependence of the Hamiltonian derives solely from the amplitude modulation, so that the preparation and free propagation of a wave packet require the knowledge of the static eigenstates. 
Thus, we begin with the description of a mathematical model system based on the combined potential of a parabolic trap and an optical lattice to exhibit main characteristics of the system\cite{Ana,Eigens,Ott}.
These theoretical features correspond closely to experimentally detected wave packets.
The experimental procedures will be thus presented in the second half, 
Sect.~\ref{ss:experiment} for making our theoretical motivation clear.

\subsection{Static Energy Spectrum}
\label{ss:spectrum}
The general Hamiltonian reads $H=-\frac{\hbar^2}{2m}\frac{\partial^2}{\partial x^2}+V_0\sin^2 (k_{r}x) + \frac{1}{2}m\omega_0^2 x^2$ where $V_0$ and $\omega_0$ denote the height of the optical lattice and the frequency of the parabolic potential, respectively. 
In this paper, we consider the following rescaled Hamiltonian
\begin{equation}
H=-\frac{\partial^2}{\partial y^2}+s \sin^2 (y) + \nu y^2,
\label{eq:re-ham}
\end{equation}
where $y$, $s$, and $\nu$ denote the rescaled position $y=k_{r}x$, the rescaled optical lattice height $s=V_0/E_r$, and the rescaled parabolic trap strength $\nu=m\omega_0^2/2E_r k_{r}^2$, respectively with $E_r=\frac{\hbar^2k_r^2}{2m}$. 
Please keep in mind that in figures to be shown later, the zero of the energy is set conveniently to the minimum eigenenergy in each case.

Let us remark in passing, several groups are working on the wave packet creation and detection; their experimental parameter values are rather similar\cite{Arlt1,Arlt2,Heinze}.
All the values in this paper are set equal to those of the Aarhus experiment\cite{Arlt1}.  The Hamburg group's values are also shown for reference.(See Table \ref{table:parameter}.)

\begin{table}[htb]
  \caption{Typical values of the parameters for Aarhus\cite{Arlt1} and Hamburg\cite{Heinze} experiments. Parameters converted for numerical analysis are also shown.
  (The Hamburg group varies $\omega_0$ and s, so that only their representative values are shown.)}
  \begin{tabular}{cccc|ccc} \hline
    \multicolumn{4}{c|}{Experimental parameters(Aarhus)} & \multicolumn{3}{|c}{Converted parameters} \\ \hline
    $\lambda$    &  $m$ ($^{87}$Rb)                     & $\omega_0$           &      $E_r$             &   $s$        &   $\nu$                    &    $J$                \\ \hline 
      914 nm     & $1.44\times 10^{-25}$ kg  & $40.6\times 2\pi$Hz  & $1.81\times 10^{-30}$ J &  16 $E_r$   & $5.51\times 10^{-5} E_r$   &  $6.06\times 10^{-3} E_r$    \\ \hline \hline
    \multicolumn{4}{c|}{Experimental parameters(Hamburg)} & \multicolumn{3}{|c}{ Converted parameters} \\ \hline
    $\lambda$    &  $m$ ($^{40}$K)                     & $\omega_0$           &      $E_r$             &   $s$        &   $\nu$                    &    $J$                \\ \hline 
      1030 nm    & $6.64\times 10^{-26}$ kg  & $50.0\times 2\pi$Hz  & $3.12\times 10^{-30}$ J &  10 $E_r$   & $2.83\times 10^{-5} E_r$   &  $2.27\times 10^{-2} E_r$    \\ \hline  
  \end{tabular}
  \label{table:parameter}
\end{table}

\begin{figure}[htbp]
 \begin{center}
 \includegraphics[width=16cm]{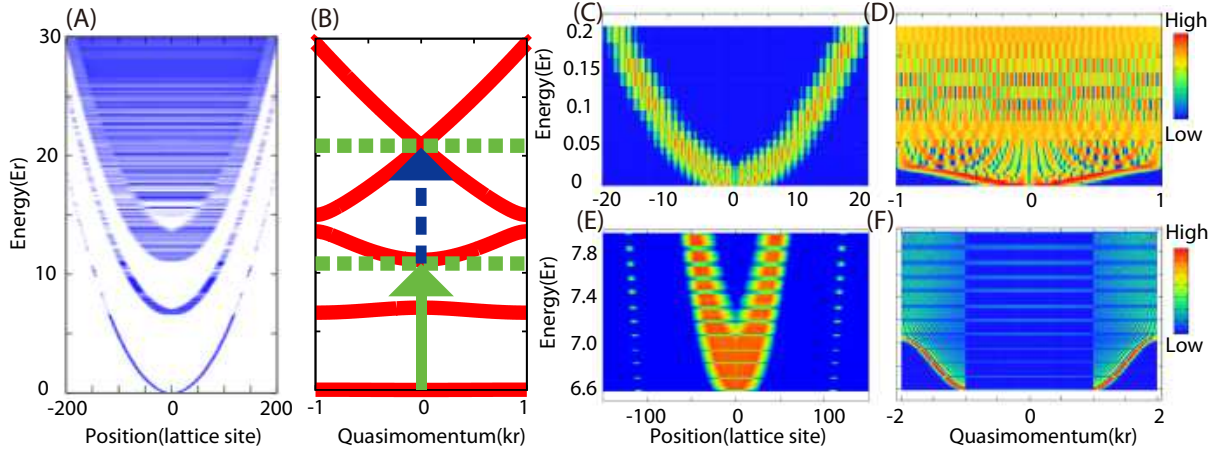}
 \end{center}
 \caption{(Color online) (A) Eigenenergies of Hamiltonian $H$ in the range of 0 to 30$Er$ represented in correlation with position space. (See text.)
 We note that the ground state energy is set to 0 in what follows.
 (B) Energy-band structure of the uniform lattice system $H_0$. 
 The energy difference at $q=0$ between ground state and 3rd bands (green arrow) and that between 3rd and 5th bands (blue arrow) serve as typical modulation frequencies in this article; specific values are 11.11$E_r$(30.4kHz) and 9.82$E_r$(26.8kHz), respectively.
 Details of the ground state band; (C) in position and (D) in quasimomentum on an expanded scale. (E) and (F) are the same for the 2nd band.
 The redder the color, the higher the density while the bluer, the lower.
 (C) Near the ground energy, the eigenstates in position space are localized around the center of the parabolic potential.
 Above the top energy of the ground-state band $E_1^1=2.42\times 10^{-2}E_r$, the eigenstate gets localized at a well-defined site away from the center of the parabolic potential.
 (D) In quasimomentum space, an eigenstate energetically near ground state gets localized at a well-defined value of $q$ and the spectrum appears similar to that of the uniform system.
 Above the top  of the ground-state band energy, the quasimomentum of eigenstates $\chi_k(y)$ spread out in the 1st Brillouin zone(BZ).
 (E) and (F) are the same as (C) and (D). At energies between $E_2^1=6.60E_r$ and $E_2^2=7.02E_r$, the structure of eigenstates resembles that of the ground band.
 And above the top energy of 2nd band, the eigenstates are localized at a well-defined place in position space, and thus cover the 2nd BZ in quasimomentum space in extended zone
representation.
 }
 \label{fig:spec}
\end{figure}

Without further ado, we present numerical results of the direct diagonalization of $H$, that is the structure of the static eigenstates including higher bands.
Fig.~\ref{fig:spec}(A) shows the eigenstates whose energies lie below 30$E_r$.
This plot needs an explanation. The probability density of each eigenfunction is shown using shade-coding as a function of coordinate $x$ while the vertical location corresponds to its eigenenergy. The darker the shade, the higher the probability density.

Because the trap potential varies very slowly as a function of $x$, it affects
the periodicity of the OL only adiabatically. The periodicity thus plays the major role in governing the dynamics of the system. 
Here we recall the concept of quasimomentum which plays an important role in our analysis. 
This quasimomentum relates to the experimentally observable momentum after an adiabatic turn-off of the trap. 
Therefore, we begin with the uniform lattice system given by
\begin{equation}
H_0=-\frac{\partial^2}{\partial y^2}+s \sin^2 (y).
\label{eq:uni-ham}
\end{equation}
 In the rest of this subsection, we interpret the eigenenergies and eigenstates of $H$ in terms of the Bloch states of $H_0$ as a function of quasimomentum.  
The trap potential will be handled shortly afterwards. 
The eigenstate corresponding to energy $E_q^n$ is the Bloch state,
\begin{equation}
\phi_q^n(y)=e^{iqy} \sum_K C_B^n (K,q) e^{2iKy},
\label{eq:q-state}
\end{equation}
where $n$, $q$ and $K \in \mathbb{Z}$ represent band index, quasimomentum, and corresponding reciprocal vector, respectively. 
The Fourier expansion coefficients $C_B^n (K,q)$ can be obtained by the Fourier series expression of Eq.~(\ref{eq:uni-ham}).
The non-zero Fourier coefficients of $s \sin^2 (y)$ corresponding only to $K=-1,0,1$, we get the recurrence formula,
\begin{equation}
(q+K)^2 C_B^n (K,q) -s C_B^n (K-1,q)/4 -s C_B^n (K+1,q)/4 =(E_q^n-s/2)C_B^n (K,q).
\label{eq:reduced-F-ham}
\end{equation}
Solving this equation, we obtain the energy bands as depicted in Fig.~\ref{fig:spec}(B) as well as the quasimomentum representation of the eigenfunction. 

As seen in Eq.~(\ref{eq:q-state}), the Bloch states are localized in quasimomentum space but delocalized in coordinate space.
The spatially localized Wannier states, on the other hand, can be built by summing over the Bloch states, namely
\begin{equation}
w_j^n(y)=\sum_q e^{iq\pi j} \phi_q^n(y),
\label{eq:w-state}
\end{equation}
where $j \in \mathbb{Z}$ is the site index representing the location in coordinate space.

Use of the Wannier states leads to the tight-binding approximation to $H$, now including the trap potential. 
The tight-binding Hamiltonian for the first band is characterized by the hopping parameter $J$ for the nearest neighboring sites,
\begin{equation}
H_{\rm TB}=-J \sum_{\langle ij\rangle} (\hat{a}_i^\dagger\hat{a}_j+\hat{a}_j^\dagger\hat{a}_i)+\nu \pi^2 \sum_j j^2 \hat{a}_{j}^\dagger\hat{a}_{j}
\label{eq:HTB}
\end{equation}
where $\hat{a}_j^\dagger$($\hat{a}_j$) is a creation(annihilation) operator, and the last term on the right pertains to the trap potential.
Let us make a momentary digression on the spectrum in this approximation as a primer to the study of the amplitude modulation in Sect.~\ref{sect:numerics}. 
 The first band is discussed in detail in Ref's~\cite{Ana,Eigens}.
 At the low energy region, eigenenergies and eigenstates of the {\it parabolic} OL Hamiltonian $H$ can be approximately given by
\begin{equation}
E_k \sim -2J +2 \sqrt{\nu \pi^2 J} (k+1/2),
\label{eq:harmonic-energy}
\end{equation}
and the corresponding eigenfunction
\begin{equation}
\chi_k (y) \sim N_k \sum_j e^{-\zeta_j^2/2} H_k (\zeta_j) w_j^1 (y),
\label{eq:harmonic-state}
\end{equation}
where $N_k$ is the normalization constant, $\zeta_j=j (J/\nu \pi^2)^{1/4}$, and $H_k$ is the $k$-th Hermite polynomials~\cite{Eigens}.
In terms of $s$, the 1st band hopping parameter and 1st band energy are approximately given\cite{Zwerger} by $J=\frac{4}{\sqrt{\pi}} s^{3/4} e^{-2\sqrt{s}}$ when $s\gg 1$ and $\pi^2 J/\nu \gg 1$.

Fig.~\ref{fig:spec}(C) and (D) show, on an expanded scale, spatial eigenstates and corresponding quasimomentum distributions, respectively, obtained by direct numerical diagonalization.
When eigenenergies lie below $4J=2.42\times 10^{-2} E_r$ (which corresponds to $E_1^1$ since we set $E_0^1=0$ for convenience), the spatial eigenstates are localized around the center of the parabolic trap potential, and the behavior of the quasimomentum distributions is similar to the Bloch band. 
On account of the trap potential which becomes increasingly steep away from the center, eigenenergy $E_k$ can exceed the top of the 1st-band, {\it i.e.}~$E_k\ge 4J$. 
The harmonic approximation with respect to $j$ thus breaks down.
In this energy region, the eigenstates are approximately given by the Wannier states $w^1_j (y)$ localized at site $j$ with energy $\nu \pi^2 j^2$.
Thus, the density distributions in quasimomentum are no longer localized and contain a wide range of quasimomentum states. (See Fig.~\ref{fig:spec}(C) and (D)).
SLEs located at the site away from the center of the parabolic potential were demonstrated in Ref.\cite{Ott}.
They suggested that SLEs play an important role in the quantum registering\cite{REGI}.
However, in the present paper, we do not access SLEs but only the quasi-free states, thus we skip detailed discussions here.

As the energy rises to the bottom of the 2nd band, namely $E_1^2=6.60E_r$ of $H_0$, the structure of the 2nd band of $H$ begins to emerge in the central part of Fig.~\ref{fig:spec}(E) (shown around -50 to 50 lattice sites) in coordinate space, resembling the 1st band.
In quasimomentum, on the other hand, the distribution is contained in $[-2,-1]$ and $[1,2]$ in the extended zone representation in (F).
Energies up to the top of the 2nd band, namely $E_0^2=7.02E_r$ of $H_0$, the spatial eigenstates are localized near the center of the parabolic potential.
And as the eigenenergy overcomes $E_0^2$, the spatial eigenstates get displaced from the center of the parabolic potential while still localized.
SLEs seen in Fig.~\ref{fig:spec}(E) around $x\simeq -100$ and $x\simeq 100$ come from the 1st band as noted earlier; the corresponding quasimomentum distributions are contained in $q\in[-1, 1]$ as in  Fig.~\ref{fig:spec}(F).

Similar structures appear up to the height of the optical lattice $s=16$, however, above this energy, the picture of the tight-binding model breaks down completely.
Here, the eigenstates are quasi-free and widely spread out in position space (Fig.\ref{fig:spec}(A)).

\subsection{Comments on experimental procedure}
\label{ss:experiment}
With the backdrop of the preceding presentation, we brief on the corresponding experimental procedure.
These experiments are initiated by superimposing the one-directional optical lattice potential adiabatically onto ultracold atoms captured in the parabolic trap.
In the Aarhus\cite{Arlt1} and Hamburg\cite{Heinze} experiments, typically about $10^5$ atoms are loaded into the parabolic OL.
After this process, the height of the optical lattice is modulated periodically\cite{Densh} in time to get the excited wave packet.
 Roughly speaking, a fraction of about $10\%$ of atoms get excited to a higher band in both experiments.
Then they follow the motion of the excited wave packet.
Experimentally, there are two techniques for detecting the motion, one is in position and the other is in quasimomentum space.
The Aarhus group hires the former way referred to as the non-destructive Faraday rotation method\cite{Arlt2}.
This method is based on the absorption imaging method, measuring the Faraday rotated components.
On the contrary, the Hamburg group hires the latter way called the band mapping technique\cite{IB}.
This method is a powerful way to observe the quasimomentum distribution by ramping down the height of the optical lattice adiabatically before the TOF imaging\cite{MG1,MKOHL,MCKAY}, {\it i.e.}~so slowly that the quasimomentum transforms itself onto the observable momentum without losing one-one correspondence.
In Sect.~\ref{sect:subsequent}, we will present numerical data, displaying the excited wave packet in both position and quasimomentum spaces.

\section{Excitation subject to amplitude modulation}
\label{sect:numerics}
Amplitude modulation of the optical lattice height has been used for exciting atoms and probing the system\cite{Rabi1,probe}.
In the Aarhus experiment\cite{Arlt1}, ultracold bosonic atoms in the ground state were transferred into the 5th band via two-photon process with a single modulation frequency, but optimization was not explored.
The primary purpose of this section is to analyze the Aarhus data on the assumption that the system can be treated as quasi-1D, but for a more noteworthy purpose, we consider how to optimize population transfer from the ground state to the 5th band with either a single frequency or two frequencies.
In accordance with the actual experiments, we assume no overlap between the two pulses in the time domain.

According to the band structure of the uniform system with $s=16$ (Fig.\ref{fig:spec}(B)), the energy difference between the 1st and the 3rd bands and that between the 3rd and the 5th bands are 11.1$E_r$ (30.4kHz) and 9.82$E_r$ (26.8kHz), respectively.
Therefore, we focus on the energy range between 9$E_r$ and 13$E_r$ and investigate the population transfer rate.
In the next subsection~\ref{ss:single-ex}, we first discuss the numerical transfer rates to the 3rd, 4th and 5th bands with a single modulation frequency.
We then analyze the direct numerical results by employing a semi-analytical method.
In subsection~\ref{ss:double}, we extend the analysis to the case of excitation by dual modulation frequencies; the first step transfers the atoms from the ground state to the 3rd band, and the second one from the 3rd band to either the 4th or 5th band.
We analyze the direct numerical result of the dual frequency excitation also by extending the semi-analytical model.

Let us note in passing that in the direct numerical calculations the Fourier Grid Hamiltonian method~\cite{FGHM} is employed for solving the eigenvalue problem and the 4th order Runge Kutta method for time propagation. 

\subsection{Excitation process with single frequency modulation}
\label{ss:single-ex}

\begin{figure}[htbp]
 \begin{center}
 \includegraphics[width=12cm]{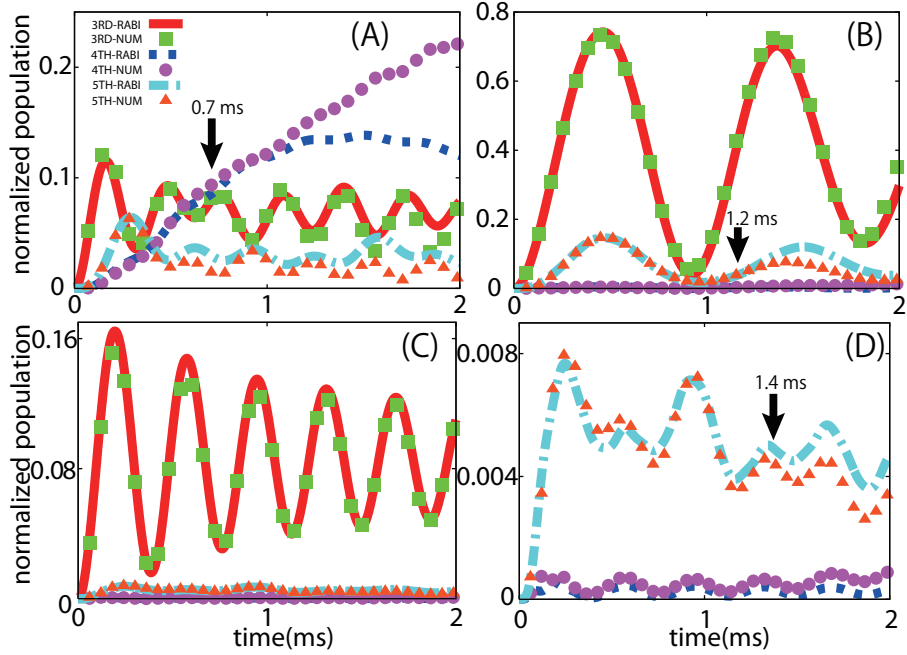}
 \end{center}
 \caption{(Color online) Band population transfer rates from the ground state as functions of pulse duration at modulation frequencies of 10.25$E_r$(28kHz) for (A), 11.35$E_r$(31kHz) for (B), and  12.08$E_r$(33kHz) for both (C) and (D).
 Red solid, blue dashed, and light-blue chain dashed curves correspond to the results of the Rabi-model for the 3rd, 4th and 5th bands, respectively.
 The discrete values represented by squares, circles, and triangles correspond to the direct numerical calculations of the 3rd, 4th and 5th bands, respectively.
 }
 \label{fig:SINGLE}
\end{figure}

There are three factors to be considered for population transfer, namely energetics, the overlap between initial and final states, and the phase evolution. We account for the energetics here.
Let us begin with transfer to the 3rd band by a single photon. The numerical results are shown with the squares in Fig.~\ref{fig:SINGLE}.
Since the excitation energy of 10.25$E_r$  falls short of the threshold energy difference 11.11$E_r$ between the ground-state and 3rd bands, the 3rd band population is much smaller than in the 11.35$E_r$ case. To be specific, observe the transfer rate to the 3rd band is about 7 times greater in Fig.~\ref{fig:SINGLE}(B)
than in Fig.~\ref{fig:SINGLE}(A).
Similarly, in the case of 12.08$E_r$, the 3rd band population is smaller than that of the 11.35$E_r$ case, being off resonance this time. Compare Fig.~\ref{fig:SINGLE}(B) and Fig.~\ref{fig:SINGLE}(C).

Next, we focus on population transfer to the 4th and 5th bands  which is realizable by two photons.
In Fig.~\ref{fig:SINGLE}(A), the 4th band population (circle in Fig.~\ref{fig:SINGLE}) is much higher than the 5th band population (triangle in Fig.~\ref{fig:SINGLE}) because the doubled frequency 20.50$E_r$ does not exceed the threshold energy difference 20.93$E_r$ between the ground-state and 5th bands.  
On the other hand, in Fig.~\ref{fig:SINGLE}(B) and (C), the 5th band population is the more dominant of the two.
Fig.~\ref{fig:SINGLE}(D), enlargement of (C), shows that the 4th band component is quite small compared to the 5th one because 24.16$E_r$ exceeds the 1st-5th energy difference, thus far off resonance.

By considering the energy diagram we could account for the trend of population transfer in a simple manner. Predicting the population transfer rate in relation to the pulse timing and peak pulse duration is a different story. In seeking a predictive method,
we find that the first order perturbation theory produces significantly small transfer rates in the situations described above. This suggests that the transfer rate under the presently considered amplitude modulation is so high that processes analogous to the multi-photon transition in quantum optics need be taken into account.
Indeed, we will use the term ``{\it multi-photon}'' to signify this analogous effect as noted.
This motivates us to introduce the following single-Q Rabi model to evaluate and interpret the transfer rate more satisfactorily.

Drawing analogies from quantum optics seems illustrative.
For instance, damping oscillations in Bloch-band population occur in ultracold atoms in an OL system as seen in several experiments \cite{Rabi1, Densh}.
Especially, fermions in a parabolic OL, the Hamburg experiment presents a clear evidence of damped oscillation in the band population~\cite{Rabi2}.
Furthermore, we also see that in the current situation there occur oscillations in relative band populations, for example in Fig.~\ref{fig:SINGLE}(B) and (C). 
We show these oscillations can be treated by a simple Rabi model.
Indeed, the Rabi model could be work well with fermionic system rather than bosonic, because the ultracold fermions could be treated as a combination of the single-particle states and the bosonic system should be take into account the effect of the atom-atom interaction.
The Rabi model with fermionic system may be discussed in an our future paper.
The damping could be caused by the LZ transition, but let us also pay attention to dephasing which causes damping-like effect in transferred populations because the n-th band population is a sum over various energy eigenstates, namely quasimomentum $q$-states. We thus separate the Rabi-oscillations and other effects.

\subsection{The single-Q Rabi model}
\label{ss:rabi-model}
Let us propose the single-Q Rabi model. The initial state being localized near the bottom, we drop the parabolic term in this model,
considering the following time-dependent perturbed Hamiltonian,
\begin{equation}
H=-\frac{\partial^2}{\partial y^2}+s \sin^2 (y) [1+\epsilon_0 \cos(\omega t)],
\label{eq:purt_ham}
\end{equation}
where $\epsilon_0$ and $\omega=\hbar \omega'/E_r$ are the modulation amplitude and rescaled modulation frequency, respectively. However, we include the effect of the parabolic trap in the analytic expression for the ground state wave function, which is
\begin{equation}
\chi_0 (y) \cong \frac{1}{ ^4 \sqrt{\pi \alpha}}  \sum_q e^{-q^2/2\alpha} \phi_q^1(y), 
\label{eq:ground-state}
\end{equation}
where $\alpha=\sqrt{\nu/\pi^2 J}$.

In this model, each $q$-component couples primarily with those components with the same $q$ value in different bands, {\it i.e.}~$\Delta q=0$ as suggested by the evaluation of the coupling matrix elements~\cite{Rabi2}.
Moreover, the coupling matrix has the following features. First, the ground state component couples strongly with only the 3rd band via one-photon process due to the energy conservation noted in the previous subsection.
On the other hand, the 3rd band component couples with the ground state, 4th, and 5th bands. Note, however, there is an additional element to consider in the process that the 1st-4th and 1st-5th couplings occur indirectly through the 3rd band.
We thus retain the ground-state, 3rd, 4th and the 5th bands under the single-Q condition $\Delta q=0$ while ignoring the 2nd band entirely.
Representing the time-dependent wave function in terms of the Bloch states as
 $\psi (y,t) = \sum_{n,q} C_Q^n(q,t) \phi_q^n(y)$, the equation of motion for the coefficient $C_Q^n(q,t)$ of a Bloch state in the $n$-th band reads
\begin{equation}
\frac{d}{dt}
\left(
\begin{array}{c}
C_Q^1 (q,t) \\
C_Q^3 (q,t) \\
C_Q^4 (q,t) \\
C_Q^5 (q,t) 
\end{array}
\right)
= -\frac{i}{2}
\left(
\begin{array}{cccc}
2\Delta_{13} & \Omega_{13} & 0 & 0\\
 \Omega_{13}  & 0  & \Omega_{34} &  \Omega_{35} \\
 0  & \Omega_{34}  &  -2\Delta_{34} &  0 \\
0  & \Omega_{35}  &  0 & -2\Delta_{35}   
\end{array}
\right)
\left(
\begin{array}{c}
C_Q^1 (q,t) \\
C_Q^3 (q,t) \\
C_Q^4 (q,t) \\
C_Q^5 (q,t) 
\end{array}
\right)
\label{eq:rabi-model}
\end{equation}
where the rotating wave approximation is employed.
Here the detuning $\Delta$ and Rabi-frequency $\Omega$ are given by
\begin{equation}
\Delta_{nm}(q,\omega')= (E_q^m - E_q^n) - \hbar \omega'
\label{eq:Delta}  
\end{equation}
and
\begin{eqnarray}
\Omega_{nm}(q) &= \epsilon_0 \left[ \sum_K C_B^m (K,q) C_B^n (K,q)(q+K)^2 \right] \\
               &= \epsilon_0 s <\phi^m_q|\sin^2(y)|\phi^n_q>.
\label{eq:rabi-freq}  
\end{eqnarray}
Note that $\hbar \omega'$ as well as $E_q^n$ are in the units of $E_r$.
The initial condition is such that $C_Q^1(q,0)=Ae^{-q^2/2\alpha}$, and the other coefficients are 0.
The normalization coefficient $A$ is $A=1/\sqrt{\sum_q |e^{-q^2/\alpha}| }$.
The $n$-th band population $B_n$ is give by $B_n(t)=\sum_{q} |C_Q^n(q,t)|^2$.

Here let us recall that this model ignores the acceleration due to the parabolic potential which would otherwise affect the single-Q condition, coupling different quasimomentum $q$ states.
The results of the direct numerical calculations and the single-Q Rabi model (Fig.~\ref{fig:SINGLE}) indeed reveal not only Rabi oscillations but also damping.

\subsection{Effects due to the gap}
\label{ss:gap}

  The populations evaluated by the two methods agree well only in a short time.
The Rabi-model breaks down after a few oscillations.
For example, the 4th band population in Fig.~\ref{fig:SINGLE}(A) by the Rabi-model departs from the direct numerical result at $t\sim 1$ms.
This moment turns out to be when the wave packet in the 5th-band crosses the boundary between 4th and 5th bands in quasimomentum space, thus entailing the LZ transition.

\begin{figure}[htbp]
 \begin{center}
 \includegraphics[width=6cm]{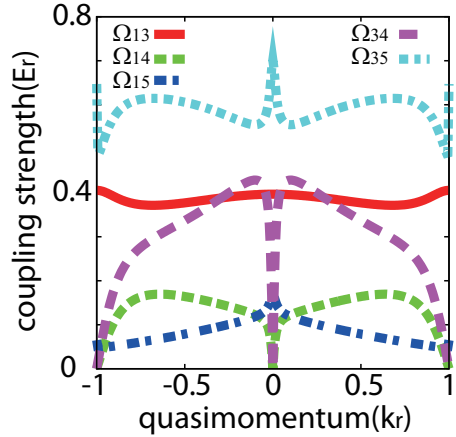}
 \end{center}
 \caption{(Color online) Coupling strength $\Omega$ for transitions 1st-3rd, 1st-4th, 1st-5th, 3rd-4th, and 3rd-5th.
 }
 \label{fig:COUPLINGS}
\end{figure}

Let us verify this observation by estimating the critical time of breakdown $t=\tau_c$ by a classical treatment~\cite{Sengstock,CT}.
Let us assume each atom obeys the canonical equations of motion based on the classical Hamiltonian
\begin{equation}
H(y,q)=E_q^n+\nu y^2. 
\label{eq:classical}  
\end{equation}
The equations of motion are thus
\begin{eqnarray}
\dot{y}(t) &= \left( \frac{\partial E_q^n}{\partial q} \right), \nonumber\\
\dot{q}(t) &= -2\nu y.
\label{eq:trajectory}  
\end{eqnarray}
which are then solved subject to appropriate initial conditions specified next.

Assume that the excited wave packet in the 5th band is initially located at $(y,q)=(0,q_{i})$ where $q_i$ is the initial value, and crosses the boundary $q=0$ at $t=\tau_c$. At this crossing, the dispersion curves are separated by a narrow band gap as depicted in Fig.~\ref{fig:spec}, thus susceptible to the LZ transition.
The value $q_{i}$ of the 5th band can be estimated simply using energy conservation which amounts to the resonance condition $\Delta_{nm}=0$ for one-photon process (transition from 3rd to 5th), and for two-photon process (transition from ground to 5th) the detuning is given by $\tilde{\Delta}_{nm}(q,\omega')=(E_q^m - E_q^n) - 2 \hbar \omega'$, so that the resonance condition reads $\tilde\Delta_{nm}=0$.

We thus obtain Table~\ref{table:critical-q} which tabulates $q_{i}$ reached by the 1st-3rd, 3rd-5th (or 3rd-4th), and 1st-5th (or 1st-4th) transitions, that is
$\Delta_{13}=0, \Delta_{34}=0, \Delta_{35}=0$
for one-photon transition, and
$\tilde{\Delta}_{15}=0, \tilde{\Delta}_{14}=0$
for two-photon transition.
Once $q_{i}$ is calculated,  the critical time $\tau_c$ is given by
\begin{equation}
\tau_c(q_{i})= \frac{1}{2\sqrt{\nu}} \int_0^{q_{i}} \frac{1}{\sqrt{E_{q_{i}}^n - E_q^n}} dq
\label{eq:critial_t}  
\end{equation}
since $E_{q_{i}}^n=E_q^n+\nu y^2$ and $dt=\frac{dq}{-2\nu y}$~\cite{Sengstock} from Eq.~(\ref{eq:trajectory}).
Let us consider as an example the specific initial value of  $q_{i}$ in the 5th band which follows readily from the information in Table.~\ref{table:critical-q} for each excitation frequency.
For 9.15 $E_r$ there are no critical value whose quasimomentum exceeds 4, due to the energy conservation, thus we could not estimate the critical time.
On the other hand, for 11.35$E_r$ and 12.08$E_r$, there are two critical quasimomentum above 4, one is given by $\Delta_{35}=0$, and the other one is $\tilde{\Delta_{15}}=0$.
The former one corresponds to the one-photon excitation process of 3rd-5th after the one-photon excitation of 1st-3rd.
The latter one corresponds to two-photon process of 1st-5th.
To estimate the critical time, we choose the smaller one.
Table~\ref{table:critical-tau} thus shows the initial values of quasimomentum $q_i$ and the corresponding values of critical time $\tau_c$.
Despite its simplicity, this estimation seems to work sufficiently well.
The role of the gap is so clear and dramatic that we shall return to this point in the next section.

\begin{table}[htb]
  \caption{Values of critical quasimomentum $q_c$ for 9.15$E_r$, 10.25$E_r$, 11.35$E_r$, and 12.08$E_r$ excitations.
  In this table, the values of $q_c$  are given in the extended representation. On the other hand, the quasimomentum $q^\prime$ is given in the reduced zone representation 
  $q^\prime=q-(n-1)$ where $n$ is the band index.
  Here only the positive values are shown.
  There is no critical value satisfying $\Delta_{13}=0$ in the cases of 9.15$E_r$ and 10.25$E_r$ excitations because these energies do not reach the threshold energy difference between 1st and 3rd bands (11.11$E_r$).}
  \begin{tabular}{|c|c|c|c|c|} \hline
    frequency            &   9.15$E_r$ &  10.25$E_r$ &  11.35$E_r$ &   12.08$E_r$  \\ \hline \hline
    $\Delta_{13}$ &   None      &  None       &  2.26       &   2.54        \\ \hline
    $\Delta_{34}$ or $\Delta_{35}$ &   3.93      &  4.07       &  4.23       &   4.35        \\ \hline
    $\Delta_{14}$ or $\Delta_{15}$ &   3.65      &  3.96       &  4.24       &   4.41        \\ \hline
  \end{tabular}
  \label{table:critical-q}
\end{table}

\begin{table}[htb]
  \caption{Values of initial quasimomentum $q_i$ and critical time $\tau_c$ for 10.25$E_r$, 11.35$E_r$, and 12.08$E_r$ excitations.
  The breakdown of the single-Q Rabi-model clearly seen in the 4th band population in Fig.~\ref{fig:SINGLE} (A) and 5th in Fig.~\ref{fig:SINGLE} (B) and (D).}
  \begin{tabular}{|c|c|c|} \hline
                &   $q_i$($k_r$)     &  $\tau_c$(ms)   \\ \hline \hline
    10.25$E_r$  &   4.07           &  0.7           \\ \hline
    11.35$E_r$  &   4.23           &  1.2           \\ \hline
    12.08$E_r$  &   4.35           &  1.4           \\ \hline
  \end{tabular}
  \label{table:critical-tau}
\end{table}

Here are some remarks concerning the restrictive single-Q condition.
We observe that the $q$ state couples strongly with the $q^\prime$ state in even bands satisfying $q^\prime=(q-1)$ when $q \geq 0$ and $q^\prime=(q+1)$ when $q \leq 0$ in the rigorous quantum calculation.
Fig.~\ref{fig:COUPLINGS} exhibits the Rabi-frequency, {\it i.e.}~coupling strength between $n$-th and $m$-th bands.
The combined effect of $\Omega_{13}$ and $\Omega_{35}$ appears much stronger than $\Omega_{15}$ alone. The two-photon 1st-5th transfer rate is thus expected to be better than the single-photon 1st-5th transfer rate. 
More detailed features of the coupling strengths as functions of quasimomentum will be discussed in the next subsection.

\subsection{Enhancement of Excitation with two-color modulation}
\label{ss:double}

\begin{figure}[htbp]
 \begin{center}
 \includegraphics[width=12cm]{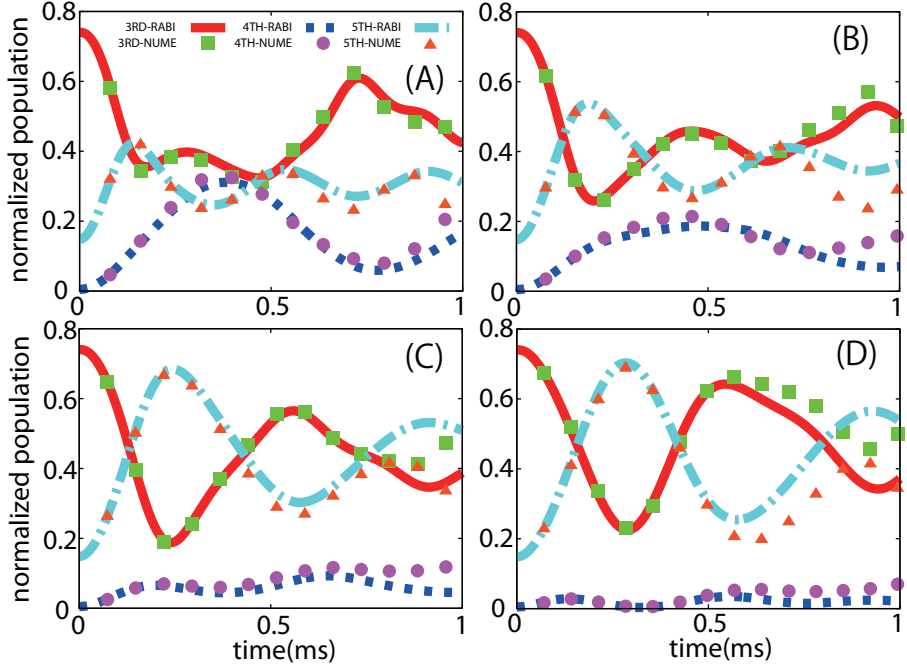}
 \end{center}
 \caption{(Color online) The $n$-th band population as a function of pulse duration with (A) 9.15$E_r$ (25kHz), (B) 9.52$E_r$ (26kHz), (C) 9.88$E_r$ (27kHz), and (D) 10.25$E_r$ (28kHz) modulations as in Fig.~\ref{fig:SINGLE}.
 In this figure, time is measured from the end of the first pulse ($t=0.45$ms in Fig.~\ref{fig:SINGLE}).
 Therefore, add 0.45ms to get the time of breakdown measured from the beginning of the first pulse.}
 \label{fig:DUAL}
\end{figure}
To achieve optimization of transfer rate, let us explore the use of dual color excitation, or two-step process.
The idea is to apply the 2nd pulse when the 1st pulse maximizes the amplitude of a desired intermediate state.
A fuller discussion requires to generalize Eq.~(\ref{eq:rabi-model}), but we shall instead show results of an educated guess as a demonstration. 
 
The Bloch coefficients $C_B^n(K,q)$ for the odd(even) band are symmetric (anti-symmetric) with respect to $K$ at $q=$0.
Therefore, at $q=$0, the coupling strength between odd and even bands are 0.
In the first process, we apply an 11.35$E_r$ excitation and stop at 0.45ms(See Fig.~\ref{fig:SINGLE}(B)).
After that we apply a pulse which has much smaller energy than 11.35$E_r$ because the energy difference between 3rd and 5th bands is much smaller than that between 1st and 3rd band.
Moreover, by using such smaller energy, we can reduce the coupling between 1st and 3rd.
According to Fig.~\ref{fig:COUPLINGS}, $\Omega_{35}$ shows strong coupling strength.
In Fig.~\ref{fig:DUAL}, we plotted results of (A) 9.15$E_r$(25kHz), (B) 9.52$E_r$(26kHz), (C) 9.88$E_r$(27kHz) and (D) 10.25$E_r$(28kHz).
Note time $t=0$ corresponds to when the first pulse modulation ends.
As might be expected, most of the 3rd band component gets transferred to the 4th or 5th band.
In all the plots, the 5th band component surpasses the 4th band component, especially in (C) and (D), most of the 3rd band component is transferred to the 5th band.

As discussed in the previous subsection \ref{ss:rabi-model}, the population in each band reflects the influence of dephasing, and the simple Rabi-model breaks down at the critical time.
In the two-color case, the critical time is derived for 11.35$E_r$(1.2ms) minus the pulse duration of the 1st pulse which equals 0.7ms. 
By comparing Fig.~\ref{fig:SINGLE}(B) and Fig.~\ref{fig:DUAL}(D), the dual color excitation makes the 5th band transfer rate three time bigger than the single color process.
This is a point worthy of special mention particularly for experiment, that the maximum transfer can be greatly enhanced by two-color excitation.

\section{Subsequent evolution in phase space}
\label{sect:subsequent}

 Time-evolution of the system represented  by $\psi(y,t)=\sum_k C_k(t) \chi_k(y)$ after amplitude modulation is straightforward as it pertains to the time-independent problem. 
Given the initial superposition coefficients $C_k (t=0)$, the amplitude of each $k$-th eigenstate of the combined potential hamiltonian evolves as
$$
C_k (t)=C_k (t=0){\rm e}^{-iE_k t}.
$$ 
We note again that $E_k$ is rescaled by $E_r$ and $t=E_rt'/\hbar$ is the rescaled time.
Despite this simplicity, the wave packet motion displays nontrivial features such as Bloch oscillation and LZ transition because the eigenstates are composed of Bloch states. 
The purpose of this section is to examine specific features closely in connection with the Aarhus experiment\cite{Arlt1}.
In their experiment, modulation is applied for the second time after a certain time of free propagation, leading to deexcitation to access the SLEs.
How much of the deexcited state becomes localized and where in space depends on the free propagation time and modulation frequency.

Apparently, the analysis of this type of process has a general merit irrespective of the bosonic or fermionic system. 
For instance, in the Hamburg experiment, free propagation of fermionic atoms is interpreted as akin to the photoconductive current.
In this section, we analyze the excitation and time-propagation of wave packets, employing both direct numerical solutions and the classical model. 

\subsection{Bloch oscillation and Landau-Zener transition}
\label{ss:Bloch-LZ}
\begin{figure}[htbp]
 \begin{center}
 \includegraphics[width=4.5cm]{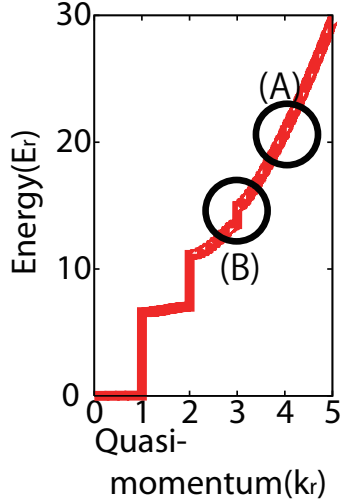}
 \end{center}
 \caption{(Color online) Band dispersion relation in the extended representation for $s=16$.
 (A) corresponds to the band gap between 4th and 5th $\sim 0.2 E_r$ which is quite small, however the LZ process can not be ignored.
 On the contrary, (B) the band gap between 3rd and 4th $\sim 1.4 E_r$ is high enough to ignore the LZ transition to 2nd band, thus we only take into account the Bragg reflection at this edge.
 }
 \label{fig:Bragg-LZ}
\end{figure}

When the wave packet evolves in time under the influence of an external linear potential, non-classical behavior occurs mostly where the band gap becomes narrow.
For instance, it is well-known that some part of the wave packet remains on the initial band while the rest makes an inter-band transition across the gap.  
This is the inter-band LZ transition. 
A good example is the gap between 4th and 5th bands at $q=0$ which is rather narrow.
To expose this point more clearly, let us show in Fig.~\ref{fig:Bragg-LZ} the energy band in extended representation. 
Region marked (A) corresponds to the narrow gap between 4th and 5th bands now located at $q=4k_r$.
At the edge of the BZ (see Region marked (B) in Fig.~\ref{fig:Bragg-LZ}), quasimomentum of the wave packet gets shifted by the lattice momentum from $q=3k_r$ to $-3k_r$, which is the Bragg reflection. 
The LZ transition may also occur at the edge.
If the loss of the Bragg reflected component is small, the wave packet is seen to oscillate in position space repeatedly, which is called the Bloch oscillation.
Thus, in the actual motion of the wave packet, the LZ transition and the Bloch oscillation occur in complicated combinations.

In the field of ultracold atomic physics, these phenomena were actually observed in early experiments in the lowest band~\cite{L-LZ,MOTION}. 
In the current study, the high-order Bragg reflection~\cite{Arlt3} and LZ process take place at the boundary of the 4th and 5th bands and at that of the 3rd and 4th bands under the influence of the parabolic potential.

\begin{figure}[htbp]
 \begin{center}
 \includegraphics[width=12cm]{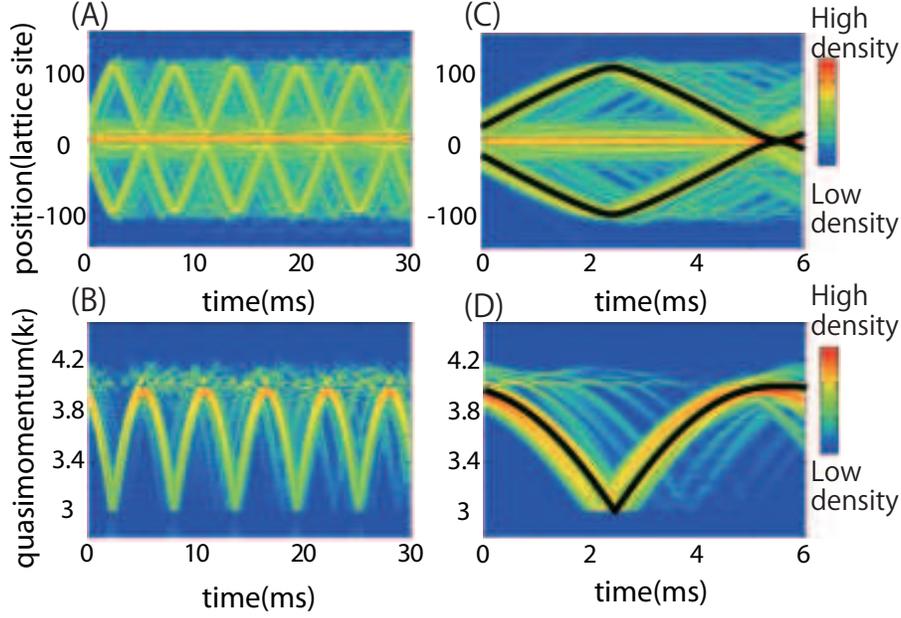}
 \end{center}
 \caption{(Color online) Free propagation of a wave packet after exerting a 10.25$E_r$ pulse, {\it i.e.}~excitation to the 4th band.
 (A): Time propagation in position space, (C) showing the first 6 ms.
 (B): Time propagation in quasimomentum space, (D) showing the first 6 ms.
The black solid curves represent most probable classical trajectories. 
 }
 \label{fig:SD-SINGLE-28}
\end{figure}

\begin{figure}[htbp]
 \begin{center}
 \includegraphics[width=12cm]{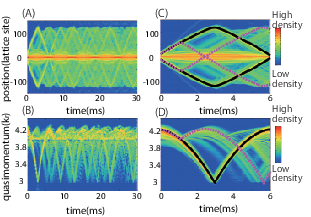}
 \end{center}
 \caption{(Color online) Free propagation after a 11.35$E_r$ pulse as in Fig.~\ref{fig:SD-SINGLE-28}, {\it i.e.}~excitation to the 5th band.
 In (C) and (D), black solid curves and pink dashed curves represent classical trajectories, the former assuming the wave packet is completely transmitted at the band gap, the latter assuming perfect reflection at the band gap.}
 \label{fig:SD-SINGLE-31}
\end{figure}

\begin{figure}[htbp]
 \begin{center}
 \includegraphics[width=12cm]{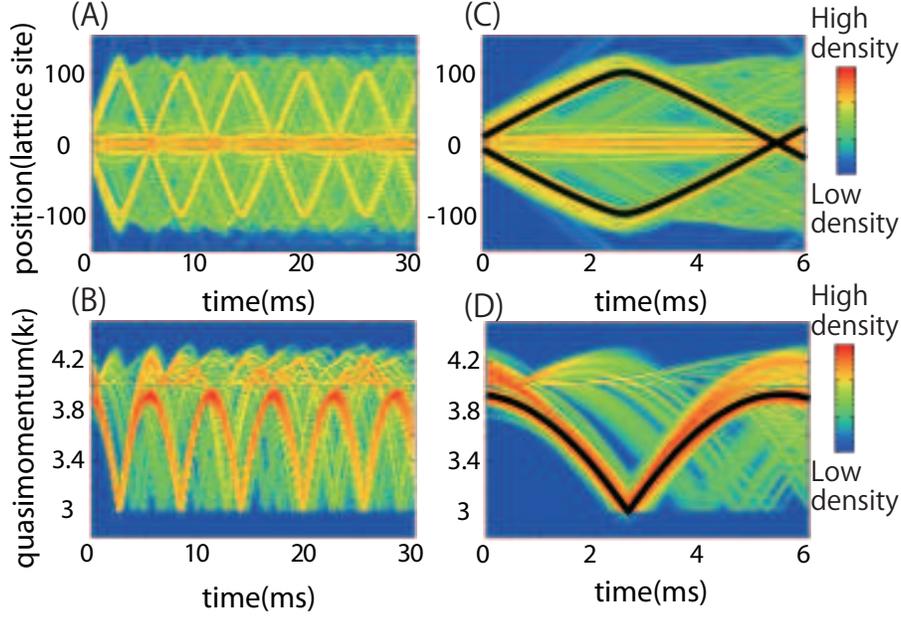}
 \end{center}
 \caption{(Color online) Free propagation after 11.35$E_r$ and 9.15$E_r$ pulses as in Fig.~\ref{fig:SD-SINGLE-28}.}
 \label{fig:SD-DUAL-25}
\end{figure}

\begin{figure}[htbp]
 \begin{center}
 \includegraphics[width=12cm]{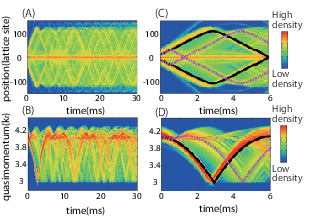}
 \end{center}
 \caption{(Color online) Free propagation after 31kHz and 28kHz pulses as in Fig.~\ref{fig:SD-SINGLE-31}.}
 \label{fig:SD-DUAL-28}
\end{figure}

Here we focus on the motion of the excited wave packet.
In Fig's~\ref{fig:SD-SINGLE-28}-\ref{fig:SD-DUAL-28}, (A) and (B) show density distributions as   functions of time for the first 30ms in position and quasimomentum spaces, respectively. (Here, time $t=0$ marks the end of the excitation pulse.)
Figures (C) and (D) are the same for the first 6ms to give fuller details of the excited wave packet motion.
The solid black lines represent the most probable classical trajectories.

First, let us account for the short-time behavior of the wave packet excited to the 4th or 5th band via absorption of $2\hbar\omega'$ regardless of whether direct two-photon excitation or sequential process through 3rd band.
The excited wave packet gets accelerated by the parabolic potential primarily along the dispersion curve.
In quasimomentum space, the excited portion of the wave packet is located near $q=0$, but spreads symmetrically to either negative or positive values of $q$.
Thus, in position space, the main component of the wave packet moves away from the center in either positive  or negative directions almost linearly in time. (See the pair of solid black curves in (C) of Fig's~\ref{fig:SD-SINGLE-28}-\ref{fig:SD-DUAL-28}.)
We show limited regions of quasimomentum space (2.8$k_r$-4.4$k_r$) in Fig.~\ref{fig:SD-SINGLE-28}-\ref{fig:SD-DUAL-28} to explain the behavior of the excited wave packet in detail.
In fact, the eigenstates pertaining to the excited wave packet contains various band components including the lower band Bloch states. 
But their amplitudes tend to be small, and the LZ process of 4th-3rd is negligibly small.
Thus, the limited region is enough for investigating the motion of the excited wave packet.
In (D) of Fig.~\ref{fig:SD-SINGLE-28}-\ref{fig:SD-DUAL-28}, the black solid curves show quadratic behavior in time due to the parabolic potential.
At a certain time, the wave packet suddenly changes its direction in position space.
It is because of the high order Bragg reflection at the boundary of the 4th and 3rd bands. The LZ transition remains rather small due to the large band gap.
In quasimomentum space, the wave packet reflected at the boundary and the positive (negative) quasimomentum component transit to negative (positive) side.
The reflected component begins to accelerate in the reverse direction and when it reaches the initial quasimomentum $|q_i|$, it reverses its direction of propagation.
This is the Bloch oscillation.

We now turn to the long-term propagation ((A) and (B) of Fig.~\ref{fig:SD-SINGLE-28}-\ref{fig:SD-DUAL-28}).
Comparing the 10.25$E_r$ and 11.35$E_r$ cases, the wave packet of the former shows more stable oscillation than the latter.
To see this, consider the LZ transition across the small gap  ($\sim 0.2E_r$) at $q=0$ between the 4th and 5th bands.
First of all, the excitation by 10.25$E_r$ modulation cannot bring the wave packet to the 5th band.
In the case of the 11.35$E_r$ excitation, the wave packet is seen to be excited at around 4.2$k_r$ in quasimomentum (NB: extended zone representation), therefore it passes the narrow band gap during free propagation.
At this occasion, part of the wave packet gets reflected back into the 5th band (dashed line in Fig.~\ref{fig:SD-SINGLE-31}(C) and (D)) and the counterpart gets transferred to the 4th band.
This LZ transition always happens when the wave packet passes the gap, thus the wave packet acquires a complicated structure in time.

Discussions along this line also apply to the dual frequency modulation case since the band structure is the same.
From the perspective of creating a stable wave packet for the purpose of interferometry, for instance, excitation to the top of the 4th band appears desirable.
On the other hand, from the perspective of exploiting a coherent bifurcation, for instance to split a wave packet, excitation to the bottom of the 5th band appears desirable.
More detailed discussions on time evolution as well as the LZ-transition are in order in the next subsection, using the classical mechanics.

\subsection{Classical Trajectories}
\label{ss:semi}
\begin{figure}[htbp]
 \begin{center}
 \includegraphics[width=6cm]{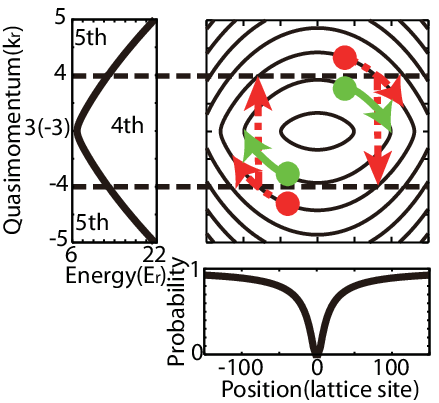}
 \end{center}
 \caption{(Color online) (A) Constant energy contours, each as a function of quasimomentum, are shown. 
 Here, quasimomentum is in extended representation thus $q=3$ corresponds to $q=-3$.
 (B) Iso-energetic lines are represented in phase space.
 The classical trajectories follow these lines under the classical hamiltonian.
 Typical initial points in 5th band (red circles) and 4th band (green circles) follow dashed arrows and solid arrows, respectively.
 When the classical point reaches the edge of 4th and 5th bands, the trajectory either goes to the other band or jumps to the other side of the same band by high-order Bragg reflection.
 (C) LZ transition probabilities between the 4th and 5th bands as functions of position (Eq.~\ref{eq:LZ-probability}) are shown.
 A fixed saddle point is located at $(y,q)=(0,4)=(0,-4)$ in this phase space representation, thus the probability at $y$=0 equals to 0. 
 }
 \label{fig:PHASE}
\end{figure}

In this subsection, the wave packet motion is treated using the classical canonical equations, Eq's~(\ref{eq:trajectory}).
Constant-energy contours in phase space are depicted in Fig.~\ref{fig:PHASE}.
Initially, the excited atoms are located at $(y,q)=(y(0),q(0))$, where an initial quasimomentum on the $n$-th band $q(0)$ is given by the energy conservation of the excitation process.
And for the initial position $y(0)$, we should take into account the migration length during the pulse. 
Here, it is simply estimated by the middle point between $(0,q(0))$ and $(y(\tau_{\rm pulse}),q(0))$, where
\begin{equation}
y(\tau_{\rm pulse})=\left( \frac{\partial E_q^n}{\partial q} \right) _{q=q(0)} \tau_{\rm pulse},
\label{eq:initial_y}  
\end{equation}
assuming quasimomentum is invariant.
Here $\tau_{\rm pulse}$ is the pulse duration. The phase space points
$(0,q(0))$ and $(y(\tau_{\rm pulse}),q(0))$ correspond to the beginning of the pulse and to the end of the pulse, respectively.
Thus the typical spatial position is given by half of $y(\tau_{\rm pulse})$.
Although strictly speaking the classical trajectory moves along the constant-energy surface satisfying the equations of motion, Eq's~(\ref{eq:trajectory}), this linear approximation works well for a short-time propagation.

Discussions on the dynamics of the excited wave packet in this subsection are based on the classical Hamiltonian, Eq.~(\ref{eq:classical}).
In the case of dual frequency excitation, we use for $y(0)$ the sum of the migration lengths of the first and second pulses.
Table~\ref{table:ini-cla} tabulates initial conditions for each case.
\begin{table}[htb]
  \caption{Values of initial position and quasimomentum for 10.25$E_r$, 11.35$E_r$, 11.35+9.15$E_r$ and 11.35+10.25$E_r$ excitations.
  Initial values of quasimomentum are estimated by the energy conservation law. See Table~\ref{table:critical-q} in subsection~\ref{ss:rabi-model}.}
  \begin{tabular}{|c|c|c|} \hline
    Initial &  y(0) &  q(0) \\ \hline \hline
    10.25$E_r$ single &  62.0& 3.96                  \\ \hline
    11.35$E_r$ single &  31.9& 4.23                  \\ \hline
    11.35+ 9.15$E_r$  &  27.4& 3.93                  \\ \hline
    11.35+10.25$E_r$  &  26.0& 4.07                  \\ \hline
  \end{tabular}
  \label{table:ini-cla}
\end{table}

We start time propagation with this set of initial conditions, and when the classical atom reaches a BZ boundary, it jumps to the other band by LZ transition or stays in the same band by Bragg reflection.
As we discussed above, we ignore the LZ process at the edge of 3rd-4th band, but include that of the 4th-5th.
Classical trajectories correspond to solid and dashed curves in Fig.~\ref{fig:SD-SINGLE-28}-\ref{fig:SD-DUAL-28}(C) and (D).
As described in the previous subsection~\ref{sect:subsequent}, a solid curve shows the trajectory which is not reflected at the 4th-5th gap, and a dashed one shows that which is reflected at the gap.
Both trajectories show a good agreement with quantum calculations. We may thus use this classical mechanical method to interpret the motion of excited atoms in both coordinate and quasimomentum spaces.

Now we focus on the dynamics at the gap of 4th and 5th bands.
The LZ transition probability is give by
\begin{equation}
P_t (n,y)=\exp \left(-\frac{\pi \delta_n^2}{16(n-1)\nu |y|} \right)
\label{eq:LZ-probability}  
\end{equation}
where $\delta_n=\Delta_n/E_r$ represents the rescaled energy gap between $n$-th and $(n-1)$-th band.
In the case of a linear external potential, acceleration does not depend on the position of the wave packet whereas in the quadratic potential, acceleration depends on position $y$.
Thus, in the system of this study, the LZ transition probability depends not only on the band index, but also on the position.
The crossing point can be determined by solving Eq's~(\ref{eq:trajectory}) with the initial condition, Eq.~(\ref{eq:initial_y}). We can thus estimate the rate.
\begin{figure}[htbp]
 \begin{center}
 \includegraphics[width=7cm]{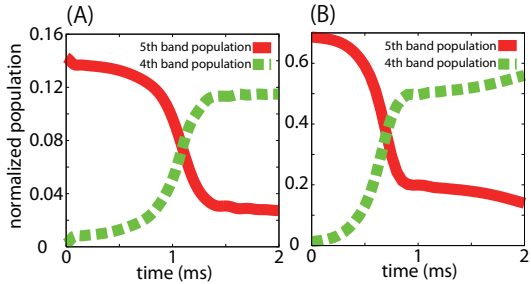}
 \end{center}
 \caption{(Color online) Time dependence of the 5th (red solid) and 4th (green dashed) band populations during free propagation.
 (A) and (B) show the results of 11.35$E_r$ single pulse and  11.35+10.25$E_r$ dual pulse case, respectively.
 In both cases, the 5th band population decreases drastically when the wave packet crosses the band gap.}
 \label{fig:LZ-result}
\end{figure}
Fig.~\ref{fig:LZ-result} shows the direct numerical result of the 4th (green dashed) and 5th (red solid) band populations as functions of time.
In both cases, 5th (4th) band population decreases (increases) drastically as the trajectory passes the gap.
Substituting the obtained position into Eq.~(\ref{eq:LZ-probability}), we evaluate the LZ probability.
Table~\ref{table:LZ-pop} shows the evaluated probabilities for 11.35$E_r$ single excitation and 11.35+10.25$E_r$ dual excitation.  The results show good agreement with Fig.~\ref{fig:LZ-result}.

\begin{table}[htb]
  \caption{Values of LZ-probability for 11.35$E_r$ single frequency excitation and 11.35+10.25$E_r$ dual frequency excitation.
   Classically estimated position $y$ and time in ms are shown to the right.}
  \begin{tabular}{|c|c|c||c|c|} \hline
    Probability & numerical & classical & position y& time(ms)\\ \hline \hline
    11.35$E_r$ single & 0.78 & 0.82  & 180& 1.11 \\ \hline
    11.35+10.25$E_r$ & 0.71 & 0.69  & 93.8& 0.58 \\ \hline
  \end{tabular}
  \label{table:LZ-pop}
\end{table}

\section{Conclusions}
\label{sect:conclusions}
In this paper, we investigated dynamics of ultracold bosonic atoms in a parabolic OL system. 
One purpose is to numerically produce coherent matter wave packets, and the other is to study their subsequent evolution in time.
 The production is experimentally effected by modulating the amplitude of the optical lattice, thus exciting the atoms coherently, and then halting its modulation for a certain period of time. The experiment then lets the wave packets propagate freely in time.
We analyzed the time dependence of the band population during the production of wave packets, and found 
 that the current experimental laser intensity is in the so-called strongly-coupled regime with atoms in an optical lattice. 
Using the language of quantum optics, we introduced the single-Q Rabi model to account for the Bloch band population. 
Simple calculations indeed showed good agreement with rigorous numerical calculations.
We also considered the two-color excitation procedure in a most straightforward way imaginable, and observed indeed enhancement in wave packet production.
 
 Energy band structures caused by the periodic optical lattice potential lead to band gaps. Here two interesting phenomena of the LZ transition and Bragg reflection occur simultaneously.
The well-known semiclassical formula for the LZ transition probability is verified in this work to yield reasonable estimates for the LZ process. 
Thus the combined effects of the Bragg reflection and the LZ transition lead to a seemingly complicated time-evolution of the wave packet. 
Such complex manifestation of wave packets suggest paradoxically that a suitable control of the optical lattice parameters may enable us to manipulate matter waves at will.

The motivated task of this paper is thus complete. 
For possible future directions, investigation of the access and the life time of SLEs\cite{Arlt1} comes immediately to our minds. 
From the theoretical viewpoint, one needs to consider both encoding and decoding of information together. 
These two tasks should proceed in tandem to check if purported aims may be fulfilled. 

The interesting issue is the effect of non-linear term because it modifies the Bloch band picture, {\it e.g.} the LZ transition probability shows asymmetric features between lowest and 1st excited band\cite{Assymetric}.
The parabolic OL system will be a good example for investigating the asymmetric LZ transition in higher band with position dependent acceleration.
Also the features of soliton solution may be investigated with the system.
In addition to the nonlinearity, the dimensionality is an interesting issue as well.
In the present work, the analysis works very well with a quasi-1D system, {\it i.e.} the tight confinement in all but the direction of the OL.
However, there is no guarantee that it works well in the case of a realistic 3D system.
Moreover, to analyze the real experiments, the experimental initial conditions need to be recreated.
This knowledge is currently incomplete, so that theoretically simulating various initial states may yield valuable information. 

The parabolic OL system is said to have a wide variety of applications, not limited to registering quantum information.
In the case of fermionic atoms, the amplitude modulation can create a particle-hole pair in analogy with photoconductivity~\cite{Heinze}.
In the Hamburg experiment~\cite{Heinze}, a variation in particle-hole recombination rate was observed as a function of the scattering length. 
Numerous other manifestations of the atom-atom interactions can be expected in the current experimental setup.
Much to be explored on the basis of the insights gained from the present linear system.

\section*{Acknowledgements}
This work was supported by JSPS KAKENHI Grant Number 26400416.
T.Y. acknowledges support from the JSPJ Institutional Program for Young Researcher Overseas Visits.
We acknowledge Dr. Alexander Itin's vital contribution to this work with his useful suggestions and guidance throughout this study.


\begin{references}
\bibitem{BECR}M. H. Anderson {\it et al.}, Science {\bf 269} 198(1995); K. B. Davis {\it et al.}, Phys. Rev. Lett. {\bf 75}, 3969(1995); C. C. Bradley {\it et al.}, Phys. Rev. Lett. {\bf 75} 1687(1995) 
\bibitem{DFGR}B. DeMarco {\it et al.}, Science {\bf 285} 1703(1999); B. DeMarco {\it et al.}, Phys. Rev. Lett. {\bf 86}, 5409(1999)
\bibitem{VORTEX}P. Vignolo, R. Fazio, and M. P. Tosi, Phys. Rev. A. {\bf 76}, 023616(2007)
\bibitem{GRAV}K. J. Hughes {\it et al.}, Phys. Rev. Lett. {\bf 102}, 150403(2009)
\bibitem{SIMU}M. Greiner {\it et al.}, Nature {\bf 415}, 39(2002)
\bibitem{GAUGE}D. Jaksch and P. Zoller, New. J. Phys. {\bf 5}, 56(2003)
\bibitem{HUM}J. Hubbard {\it et al.}, Proc. R. Soc. Lond. A. {\bf 276}, vol.1365, 238 (1963)
\bibitem{BHUM}H. A. Gersch and G. C. Knollman, Phys. Rev. {\bf 129}, 959(1963)
\bibitem{LBO}O. Morsch {\it et al.}, Phys. Rev. Lett. {\bf 87}, 140402(2001)
\bibitem{LZZ}M. Cristiani {\it et al.}, Phys. Rev. A. {\bf 65}, 063612(2002)
\bibitem{Ana}A. M. Rey, G. Puppilo, C. W. Clarck, and C. J. Williams, Phys. Rev. A. {\bf 72}, 033616(2005)
\bibitem{Eigens}M. Valiente and D. Petrosyan, Eur. Phys. Lett. {\bf 83}, 30007(2008)
\bibitem{Single}C. Hooley, and J. Quintanilla, Phys. Rev. Lett. {\bf 93}, 080404(2004)
\bibitem{Ott}H. Ott {\it et al.}, Phys. Rev. A. {\bf 93}, 120407(2004)
\bibitem{REGI}L. Viverit, C. Menotti, T. Calarco, and A. Smerzi, Phys. Rev. Lett. {\bf 93}, 110401 (2004)
\bibitem{Arlt1}J. F. Sherson {\it et al.}, New J. Phys. {\bf 14}, 083013 (2012), P. L. Pedersen {\it et al.}, Phys. Rev. A. {\bf 88}, 023620 (2013).
\bibitem{Heinze}J. Heinze {\it et al.}, Phys. Rev. Lett. {\bf 110}, 085302(2013)
\bibitem{Arlt2}M. Gajdacz {\it et al.}, Rev. Sci. Instrum. {\bf 84}, 083105(2013)
\bibitem{Chin}C. Chin {\it et al.}, Rev. Mod. Phys. {\bf 82}, 1225(2010)
\bibitem{Zwerger}W. Zwerger, J. Opt. B: Quantum Semiclassical Opt {\bf 5}, S9(2003)
\bibitem{Densh}J. H. Denschlag {\it et al.}, J. Phys. B. {\bf 35}, 3095(2002)
\bibitem{IB}I. Bloch, J. Dalibard, and W. Zwerger, Rev. Mod. Phys. {\bf 80}, 88(2008)
\bibitem{MG1}M. Greiner {\it et al.}, PRL {\bf 87}, 160405(2001)
\bibitem{MKOHL}M. K\"ohl {\it et al.}, PRL {\bf 94}, 080403(2005)
\bibitem{MCKAY}D. McKay, M. White, and B DeMarco, Phys. Rev. A. {\bf 79}, 063605(2009)
\bibitem{Rabi1}M. C. Fischer {\it et al.}, Phys. Rev. A. {\bf 58}, R2648(1998)
\bibitem{probe}T. St\"oferle {\it et. al.}, Phys. Rev. Lett. {\bf 92}, 130403 (2004); D. Greif {\it et. al.}, Phys. Rev. Lett. {\bf 106}, 145302 (2011)
\bibitem{FGHM}C. C. Marston and G. G. Balint-Kurti, J. Chem. Phys. {\bf 91}, 3571 (1989)
\bibitem{Rabi2}B. Hundt, Diploma thesis, Universit\"at Hamburg (2011)
\bibitem{Sengstock}J. Heinze {\it et al.}, Phys. Rev. Lett. {\bf 107}, 135303(2011)
\bibitem{CT}J. Bland and A. R. Kolovsky, Eur. Phys. J. D {\bf 41}, vol 2. 331(2006)
\bibitem{L-LZ}M. B. Dahan {\it et al.}, Phys. Rev. Lett. {\bf 76}, 4508(1996);B. P. Anderson and M. A. Kasevich, Science {\bf 282}, 1686(1998)
\bibitem{MOTION}F. S. Cataliotti{\it et al.}, Science {\bf 293}, 843(2001);T. Anker et al, Opt. Express {\bf 12}, 11(2004)
\bibitem{Arlt3}S. J. Park {\it et al.}, Phys. Rev. A. {\bf 85}, 033602 (2012)
\bibitem{Assymetric}M. Jona-Lasinio {\it et al.}, Phy. Rev. Lett. {\bf 91}, 230406 (2003)

\end{references}
\end{document}